\documentclass[aps,pra,reprint,onecolumn,notitlepage,showkeys,showpacs,amsfonts,
               amsmath]{revtex4-1}
\usepackage{amsthm}
\usepackage{enumerate}

\theoremstyle{definition}

\newtheorem{remark}{Remark}[section]

\newcommand{\Eqref}[1]{Equation~\eqref{#1}}

\newcommand{\secref}[1]{Section~\ref{#1}}

\newcommand{\Secref}[1]{Section~\ref{#1}}

\newcommand{\dd}{\mathrm{d}}

\newcommand{\ud}{\,\dd}
\newcommand{\nbr}[1]{$#1$\nobreakdash-\hspace{0pt}}
\providecommand{\abs}[1]{\lvert#1\rvert}

\DeclareMathOperator{\id}{id}

\newcommand{\iftwocolumn}[2]{#2}

\numberwithin{equation}{section}

\begin{document}

\title{Natural star-products on symplectic manifolds and related quantum
       mechanical operators}
\author{Maciej B{\l}aszak}
\email[Electronic address: ]{blaszakm@amu.edu.pl}
\author{Ziemowit Doma{\'n}ski}
\email[Electronic address: ]{ziemowit@amu.edu.pl}
\affiliation{Faculty of Physics, Adam Mickiewicz University\\
             Umultowska 85, 61-614 Pozna{\'n}, Poland}
\date{\today}

\begin{abstract}
In this paper is considered a problem of defining natural star-products on
symplectic manifolds, admissible for quantization of classical Hamiltonian
systems. First, a construction of a star-product on a cotangent bundle to an
Euclidean configuration space is given with the use of a sequence of pair-wise
commuting vector fields. The connection with a covariant representation of such
a star-product is also presented. Then, an extension of the construction to
symplectic manifolds over flat and non-flat pseudo-Riemannian configuration
spaces is discussed. Finally, a coordinate free construction of related quantum
mechanical operators from Hilbert space over respective configuration space is
presented.
\end{abstract}



\maketitle

\section{Introduction}
\label{sec:1}
The formalism of quantization of systems described by configuration spaces
in the form of Euclidean spaces is well established and confirmed by
experiments. The next step should be theory of quantization of systems defined
on curved spaces, e.g. systems with constraints or systems coupled with
classical gravitational fields. This task however constitutes some problems as,
because of the lack of experiments, it is difficult to find a proper
generalization of the quantization formalism. The only thing one can do is to
work on the mathematical level and try to find some distinguished quantization
schemes with interesting properties from the wast number of possibilities.

This paper aims in a discussion of this problem from a point of view of
deformation quantization theory. In this approach to quantum mechanics the
quantization is basically given by introducing a star-product on a phase space.
Thus in this paper we will deal first with a problem of defining natural
star-products on symplectic manifolds (phase spaces), and second with their
appropriate operator representation in a Hilbert space over configuration space.

In the work of \citet{Bayen:1978a,Bayen:1978b} there was presented a
construction of a star-product on a symplectic manifold endowed with a flat
symplectic linear connection. Later \citet{Fedosov:1994} presented a
construction of an admissible star-product for a general symplectic connection.
The resulting star-products were given in a covariant form independent on the
coordinate system. This results, although elegant, are difficult to use in
computations. In this paper first we discuss an alternative way of introducing
a star-product. It is base on a
definition of a star-product with the use of a sequence of pair-wise commuting
vector fields defined on a symplectic manifold. In this way equations for
star-products are of simpler form and can be easier used in computations.
Moreover, we discuss the connection of the vector field representation of the
star-product to the covariant form of the star-product (\secref{sec:2}).

An important property of the star-product is an equivalence with the Moyal
product. This allows introduction of the operator approach to quantum mechanics
\cite{Blaszak:2013,Blaszak:2013b}. It is known how to pass to the operator
representation of quantum mechanics in the case of Euclidean configuration
spaces. In a general case we can use the fact that for any classical and quantum
canonical coordinate system the star-product is equivalent with the Moyal
product. This property allows to construct the operator representation of
quantum mechanics from the knowledge of this construction in Euclidean case.
In \secref{sec:4} we construct the equivalence for the star-product written in
a covariant form on a flat symplectic manifold.

In \secref{sec:5} we discuss how to introduce star-products on a general
symplectic manifold in a natural manner. We also present an example of such
products, which construction involves symplectic linear connection on a
symplectic manifold.

\Secref{sec:6} is devoted to a problem of associating to star-algebras certain
algebras of operators defined on particular Hilbert spaces. Usually, in the
literature, one can find this connection for a Moyal star-product written in
Cartesian coordinates. The general case seems not to be considered yet.
We describe a connection between star-algebras and respective operator algebras
for a very general family of star-products considered in the paper. In
particular we describe a procedure of associating, in a coordinate independent
way, to every phase space function an operator defined on a Hilbert space of
square integrable functions defined on a configuration space. We also give
examples of operators linear, quadratic and cubic in momenta written in an
invariant form and derived for a very general star-product defined on a
symplectic manifold over a curved pseudo-Riemannian space.

In \secref{sec:7} are made some remarks about quantization of classical
Hamiltonian systems. We also discuss a problem, using the results presented in
the paper, of choosing a physically admissible quantizations for Hamiltonian
systems from phase spaces considered in the paper.

\section{The case of a symplectic manifold $T^*E^N$}
\label{sec:2}
Let us consider an \nbr{N}dimensional Euclidean space $E^N$. The cotangent
bundle $T^*E^N$ to this space is an \nbr{2N}dimensional manifold naturally
endowed with a symplectic structure $\omega$. Let us choose some Euclidean
coordinate system $(x^1,\dotsc,x^N)$ on $E^N$. We can extend this coordinate
system to a canonical (Darboux) coordinate system
$(x^1,\dotsc,x^N,p_1,\dotsc,p_N)$ on $T^*E^N$, which we will call an Euclidean
coordinate system on the symplectic manifold $T^*E^N$. In this coordinates
the symplectic form $\omega$ takes the form $\dd{p_i} \wedge \dd{x^i}$. Also
the Poisson tensor $\mathcal{P} = \omega^{-1}$ related to the symplectic form
$\omega$ can be written in the form
\begin{equation}
\mathcal{P} = \partial_{x^i} \wedge \partial_{p_i}.
\label{eq:31}
\end{equation}
\Eqref{eq:31} shows that the Poisson tensor $\mathcal{P}$ can be decomposed into
a wedge product of pair-wise commuting vector fields. However, such
decomposition is not unique. There are different sets of commuting
vector fields $X_1,\dotsc,X_N,Y_1,\dotsc,Y_N$ such that
\begin{equation}
\mathcal{P} = \sum_{i=1}^N X_i \wedge Y_i.
\label{eq:16}
\end{equation}

In what follows we will define a family of star-products on the symplectic
manifold $T^*E^N$. Let $X_i,Y_i$ be a sequence of
pair-wise commuting global vector fields from the decomposition \eqref{eq:16} of
the Poisson tensor $\mathcal{P}$. Define a star-product by the formula
\begin{equation}
f \star g = f \exp \left( \frac{1}{2} i\hbar \sum_i \overleftarrow{X_i}
    \overrightarrow{Y_i} - \frac{1}{2} i\hbar \sum_i \overleftarrow{Y_i}
    \overrightarrow{X_i} \right) g.
\label{eq:17}
\end{equation}
From the commutativity of vector fields $X_i,Y_i$ follows the associativity of
the star-product. As was pointed out earlier the sequence $X_i,Y_i$ is
not uniquely specified by the Poisson tensor, thus we can define the whole
family of star-products related to the same Poisson tensor.

For a given sequence of vector fields $X_i,Y_i$ from the decomposition
\eqref{eq:16} of the Poisson tensor $\mathcal{P}$ there exists a global
coordinate system $(x,p)$ in which $X_i,Y_i$ are coordinate vector fields, i.e.
$X_i = \partial_{x^i}$, $Y_i = \partial_{p_i}$. Such coordinate system is of
course a Darboux coordinate system associated with the symplectic form $\omega$.
In this coordinates the star-product \eqref{eq:17} takes the form of
a product
\begin{equation}
f \star g = f \exp \left( \frac{1}{2} i\hbar \overleftarrow{\partial_{x^i}}
    \overrightarrow{\partial_{p_i}}
    - \frac{1}{2} i\hbar \overleftarrow{\partial_{p_i}}
    \overrightarrow{\partial_{x^i}} \right) g,
\label{eq:1}
\end{equation}
which is called a Moyal product \cite{Moyal:1949,Bayen:1978a,Bayen:1978b}. The
coordinate system $(x,p)$ we will call the natural coordinate system of the
star-product.

The structure of the symplectic manifold $T^*E^N$ distinguishes one product
from the presented family of star-products, namely the one for which the natural
coordinate system is the Euclidean coordinate system. Such star-product is
indeed uniquely defined since coordinate vector fields of Euclidean coordinate
systems are related to each other by linear symplectic transformations and
such transformations do not change the star-product \eqref{eq:17}. This
distinguished star-product will be called a canonical star-product on $T^*E^N$.

In what follows let us write the canonical star-product on $T^*E^N$ in a
different form. To do this let us first write it in a Darboux coordinate system
induced from an arbitrary curvilinear coordinates on $E^N$. Let
$\phi \colon (x'^1,\dotsc,x'^N) \mapsto (x^1,\dotsc,x^N)$ be a change of
coordinates from arbitrary curvilinear coordinates $(x'^1,\dotsc,x'^N)$ to
Euclidean coordinates $(x^1,\dotsc,x^N)$. The transformation $\phi$ on $E^N$
induces a canonical transformation $(x',p') \mapsto T(x',p') = (x,p)$ on the
symplectic manifold $T^*E^N$:
\begin{align*}
x^i & = \phi^i(x'), \\
p_i & = [(\phi'(x'))^{-1}]^j_i p'_j,
\end{align*}
where $[(\phi'(x'))^{-1}]^j_i$ denotes an inverse matrix to the Jacobian matrix
$[\phi'(x')]^i_j = \frac{\partial \phi^i}{\partial x^j}(x')$ of $\phi$. The
transformation $T$ is called a point transformation.

The canonical star-product in Euclidean coordinates takes the form of a Moyal
product \eqref{eq:1}. The Moyal product \eqref{eq:1} under the point
transformation $T$ transforms to the following star-product:
\begin{equation}
f \star^{(x',p')} g = f \exp \left( \frac{1}{2} i\hbar \overleftarrow{D_{x'^i}}
    \overrightarrow{D_{p'_i}} - \frac{1}{2} i\hbar \overleftarrow{D_{p'_i}}
    \overrightarrow{D_{x'^i}} \right) g,
\label{eq:9}
\end{equation}
where
\begin{align*}
D_{x'^i} & = [(\phi'(x'))^{-1}]^j_i \left( \partial_{x'^j}
    + \Gamma^r_{jl}(x') p'_r \partial_{p'_l} \right), \\
D_{p'_i} & = [\phi'(x')]^i_j \partial_{p'_j}
\end{align*}
is a transformation of Euclidean coordinate vector fields $\partial_{x^i}$,
$\partial_{p_i}$ to a new coordinate chart,
and $\Gamma^i_{jk}(x') = [(\phi'(x'))^{-1}]^i_r [\phi''(x')]^r_{jk}$
($[\phi''(x')]^i_{jk} = \frac{\partial^2 \phi^i}{\partial_{x'^j}
\partial_{x'^k}}(x')$ is the Hessian of $\phi$). Note that the symbols
$\Gamma^i_{jk}(x')$ are the Christoffel symbols for the $(x'^1,\dotsc,x'^N)$
coordinates, associated to the standard linear connection $\nabla$ on the
configuration space $E^N$. Formula \eqref{eq:9} can be written in the form
\begin{widetext}
\begin{equation}
f \star^{(x',p')} g = \sum_{n,m=0}^\infty \frac{1}{n!m!} (-1)^m
    \left(\frac{i\hbar}{2}\right)^{n+m}
    (D^{j_1 \dotsc j_m}_{i_1 \dotsc i_n} f)
    (D^{i_1 \dotsc i_n}_{j_1 \dotsc j_m} g),
\label{eq:10}
\end{equation}
where operators $D^{j_1 \dotsc j_m}_{i_1 \dotsc i_n}$ are given recursively by
\begin{subequations}
\label{eq:12}
\begin{align}
D^{j_1 \dotsc j_m}_{i_1 \dotsc i_{n+1}}f & =
    D_{i_{n+1}}(D^{j_1 \dotsc j_m}_{i_1 \dotsc i_n}f)
    - \Gamma^k_{i_1 i_{n+1}} D^{j_1 \dotsc j_m}_{k \dotsc i_n}f - \dotsb
    - \Gamma^k_{i_n i_{n+1}} D^{j_1 \dotsc j_m}_{i_1 \dotsc k}f \nonumber \\
& \quad {} + \Gamma^{j_1}_{k i_{n+1}} D^{k \dotsc j_m}_{i_1 \dotsc i_n}f +\dotsb
    + \Gamma^{j_m}_{k i_{n+1}} D^{j_1 \dotsc k}_{i_1 \dotsc i_n}f,
\label{eq:12a} \\
D^{j_1 \dotsc j_{m+1}}_{i_1 \dotsc i_n}f & =
    D^{j_{m+1}}(D^{j_1 \dotsc j_m}_{i_1 \dotsc i_n}f),
\label{eq:12b} \\
D_i f & = \partial_{x'^i}f + \Gamma^k_{ij} p'_k \partial_{p'_j}f,
\label{eq:12c} \\
D^j f & = \partial_{p'_j}f,
\label{eq:12d}
\end{align}
\end{subequations}
\end{widetext}
where $\{D_i,D^j\}$ is a so called adopted frame on $T^*E^N$ \cite{Mok:1977}.
Note that the upper indices in the operator $D^{j_1\dotsc j_m}_{i_1\dotsc i_n}$
commute with the lower indices, i.e. it does not matter if, when calculating
$D^{j_1 \dotsc j_m}_{i_1 \dotsc i_n}f$, we first use formula \eqref{eq:12a} and
then \eqref{eq:12b} or vice verse.

Equation \eqref{eq:10} can be written in the form
\begin{widetext}
\begin{equation}
f \star^{(x',p')} g = \sum_{k=0}^\infty \frac{1}{k!}
    \left(\frac{i\hbar}{2}\right)^k \sum_{n=0}^k \binom{k}{n} (-1)^{k-n}
    (\underbrace{\tilde{\nabla} \dotsm \tilde{\nabla}}_k f)_{i_1 \dotsc i_n
    \bar{j}_1 \dotsc \bar{j}_{k-n}}
    (\underbrace{\tilde{\nabla} \dotsm \tilde{\nabla}}_k g)_{\bar{i}_1 \dotsc
    \bar{i}_n j_1 \dotsc j_{k-n}},
\label{eq:13}
\end{equation}
\end{widetext}
where $\bar{i} = N + i$ and $\tilde{\nabla}$ is a linear connection on the
symplectic manifold $T^*E^N$, which components in the frame $\{D_i,D^j\}$ are
equal
\begin{equation*}
\tilde{\Gamma}^i_{jk} = \Gamma^i_{jk}, \quad
\tilde{\Gamma}^{\bar{i}}_{\bar{j} k} = -\Gamma^j_{ik}
\end{equation*}
with the remaining components equal zero. Equation \eqref{eq:13} can be written
in the form
\begin{widetext}
\begin{equation}
f \star^{(x',p')} g = \sum_{k=0}^\infty \frac{1}{k!}
    \left(\frac{i\hbar}{2}\right)^k
    \sum_{n=0}^k \binom{k}{n} A^{\mu_1 \nu_1} \dotsm A^{\mu_n \nu_n}
    B^{\mu_{n+1} \nu_{n+1}} \dotsm B^{\mu_k \nu_k}
    (\underbrace{\tilde{\nabla} \dotsm \tilde{\nabla}}_k f)_{\mu_1\dotsc \mu_k}
    (\underbrace{\tilde{\nabla} \dotsm \tilde{\nabla}}_k g)_{\nu_1\dotsc \nu_k},
\label{eq:14}
\end{equation}
where
\begin{equation*}
A = \begin{pmatrix} 0 & I \\ 0 & 0 \end{pmatrix}, \quad
B = \begin{pmatrix} 0 & 0 \\ -I & 0 \end{pmatrix}.
\end{equation*}
Equation \eqref{eq:14} takes the form
\begin{align}
f \star^{(x',p')} g & = \sum_{k=0}^\infty \frac{1}{k!}
    \left(\frac{i\hbar}{2}\right)^k
    (A + B)^{\mu_1 \nu_1} \dotsm (A + B)^{\mu_k \nu_k}
    (\underbrace{\tilde{\nabla}\dotsm\tilde{\nabla}}_k f)_{\mu_1\dotsc\mu_k}
    (\underbrace{\tilde{\nabla}\dotsm\tilde{\nabla}}_k g)_{\nu_1\dotsc\nu_k}
    \nonumber \\
& = \sum_{k=0}^\infty \frac{1}{k!} \left(\frac{i\hbar}{2}\right)^k
    \omega^{\mu_1 \nu_1} \dotsm \omega^{\mu_k \nu_k}
    (\underbrace{\tilde{\nabla} \dotsm \tilde{\nabla}}_k f)_{\mu_1\dotsc \mu_k}
    (\underbrace{\tilde{\nabla} \dotsm \tilde{\nabla}}_k g)_{\nu_1\dotsc \nu_k},
\label{eq:15}
\end{align}
\end{widetext}
where
\begin{equation*}
\omega = A + B = \begin{pmatrix} 0 & I \\ -I & 0 \end{pmatrix}.
\end{equation*}
Since $D_i \wedge D^j = \partial_{x'^i} \wedge \partial_{p'_j}$,
$\omega^{\mu \nu}$ are components of the Poisson tensor in the Darboux frame
$\{\partial_{x'^i},\partial_{p'_j}\}$ as well as in the adopted frame
$\{D_i,D^j\}$.

The Christoffel symbols of the linear connection $\tilde{\nabla}$ in the Darboux
coordinate frame take the form
\begin{gather}
\tilde{\Gamma}^i_{jk} = \Gamma^i_{jk}, \quad
\tilde{\Gamma}^{\bar{i}}_{\bar{j} k} = -\Gamma^j_{ik}, \quad
\tilde{\Gamma}^{\bar{i}}_{j \bar{k}} = -\Gamma^k_{ji},
\iftwocolumn{\nonumber \\}{\quad}
\tilde{\Gamma}^{\bar{i}}_{jk} = p_l(\Gamma^r_{jk} \Gamma^l_{ri}
    + \Gamma^r_{ik} \Gamma^l_{rj} - \Gamma^l_{ij,k}),
\label{eq:32}
\end{gather}
with the remaining components equal zero. It is straightforward to check that
$\tilde{\nabla}$ is symplectic, i.e. $\tilde{\nabla}\omega = 0$. Moreover, from
flatness of the configuration space $E^N$ follows that $\tilde{\nabla}$ is flat
and torsionless.

Thus we wrote the canonical star-product on $T^*E^N$ in a covariant form
\eqref{eq:15}, where $\tilde{\nabla}$ is a connection induced from a standard
Levi-Civita connection on $E^N$. Other star-products on $E^N$ also can be
written in a covariant form \eqref{eq:15}. As a linear connection
$\tilde{\nabla}$ one has to take a connection which components in a natural
coordinate system vanish. However, such connection is not related to a standard
Levi-Civita connection on $E^N$.

\Eqref{eq:32} defines a lift of the Levi-Civita connection on $E^N$ to a
symplectic connection on $T^*E^N$. It is possible to define a lift of the
Levi-Civita connection $\Gamma^i_{jk}$ on a general Riemannian manifold
$\mathcal{Q}$ to a symplectic and torsionless connection
$\tilde{\Gamma}^\alpha_{\beta \gamma}$ on the cotengent bundle $T^*Q$.
The resulting connection in the Darboux coordinate frame is given by the
formulas
\begin{gather}
\tilde{\Gamma}^i_{jk} = \Gamma^i_{jk}, \quad
\tilde{\Gamma}^{\bar{i}}_{\bar{j} k} = -\Gamma^j_{ik}, \quad
\tilde{\Gamma}^{\bar{i}}_{j \bar{k}} = -\Gamma^k_{ji},
\iftwocolumn{\nonumber \\}{\quad}
\tilde{\Gamma}^{\bar{i}}_{jk} = p_l(\Gamma^r_{jk} \Gamma^l_{ri}
    + \Gamma^r_{ik} \Gamma^l_{rj} - \Gamma^l_{ij,k} - \tfrac{1}{3}R^l_{ijk}
    - \tfrac{1}{3}R^l_{jik}),
\label{eq:33}
\end{gather}
with the remaining components equal zero. In the adopted frame $\{D_i,D^j\}$ the
connection $\tilde{\Gamma}^\alpha_{\beta \gamma}$ takes the form
\begin{gather}
\tilde{\Gamma}^i_{jk} = \Gamma^i_{jk}, \quad
\tilde{\Gamma}^{\bar{i}}_{\bar{j} k} = -\Gamma^j_{ik},
\iftwocolumn{\nonumber \\}{\quad}
\tilde{\Gamma}^{\bar{i}}_{jk} = -\frac{1}{3}p_l(R^l_{ijk} + R^l_{jik}),
\label{eq:34}
\end{gather}
with the remaining components equal zero. As we will see later on a symplectic
manifold endowed with a symplectic torsionless connection it is possible to
distinguish a star-product. In the majority of physically interesting cases as
the symplectic manifold is taken the cotangent bundle to a configuration space
being a Riemannian manifold. In such case there exists a distinguished
connection and thus a star-product which can be used to introduce quantization.
More about lifts of connections can be found in \cite{Plebanski:2001,Mok:1977}.

\section{The case of a symplectic manifold $T^*\mathcal{Q}$ with a flat base
manifold $\mathcal{Q}$}
\label{sec:4}
The star-product \eqref{eq:17} can be defined on more general symplectic
manifolds. Let $\mathcal{Q}$ be an \nbr{N}dimensional flat pseudo-Riemannian
manifold with a property that every two points of $\mathcal{Q}$ can be connected
by exactly one geodesic. On such manifold there exists a global Riemann normal
coordinate system $(x^1,\dotsc,x^N)$. Every such coordinate system is
parametrized by a point $x \in \mathcal{Q}$ and a basis $e_1,\dotsc,e_N$ in
$T_x\mathcal{Q}$. Using the flatness of the manifold $\mathcal{Q}$ one can check
that Riemann normal coordinate systems transform according to the rule
\begin{equation}
x'^i = A^i_j x^j + x_0^i,
\label{eq:35}
\end{equation}
where $x_0^i$ are the coordinates of the origin of the second coordinate system
from the perspective of the first coordinate system, and $A^i_j$ is a matrix
transforming the basis $e_1,\dotsc,e_N$ of the first coordinate system to a
parallel transported basis $e'_1,\dotsc,e'_N$ of the second coordinate system.

The Riemann normal coordinate system $(x^1,\dotsc,x^N)$ induces a global
canonical coordinate system $(x^1,\dotsc,x^N,\allowbreak p_1,\dotsc,p_N)$ on a
symplectic manifold $T^*\mathcal{Q}$. We will call this coordinate system a
Riemann normal coordinate system on $T^*\mathcal{Q}$. The canonical Poisson
tensor $\mathcal{P}$ on $T^*\mathcal{Q}$ using the Riemann normal coordinates
can be globally written in the form \eqref{eq:31}.

Using the coordinate vector fields of the Riemann normal coordinate system on
$T^*\mathcal{Q}$ we can introduce a star-product on the symplectic manifold
$T^*\mathcal{Q}$ by the formula \eqref{eq:1}. The Riemann normal coordinate
system is then a natural coordinate system for this star-product. Such
star-product is independent on the choice of the Riemann normal coordinate
system since, in accordance to \eqref{eq:35}, coordinate vector fields of
Riemann normal coordinate systems are related to each other by linear symplectic
transformations and such transformations do not change the star-product.
Thus on the symplectic manifold $T^*\mathcal{Q}$ there is a distinguished
star-product from the family of star-products \eqref{eq:17} given by the
decompositions \eqref{eq:16} of the Poisson tensor. We will call this product a
canonical star-product on $T^*\mathcal{Q}$.

For Riemann normal coordinates the Christoffel symbols $\Gamma^i_{jk}$ of the
Levi-Civita connection $\nabla$ on $\mathcal{Q}$ vanish. Thus also vanish the
Christoffel symbols $\tilde{\Gamma}^\alpha_{\beta \gamma}$ of the lift
\eqref{eq:33} of the connection $\nabla$ to a connection $\tilde{\nabla}$ on
$T^*\mathcal{Q}$. This shows that the canonical star-product on $T^*\mathcal{Q}$
can be written in a covariant form
\begin{align}
f \star g & = \sum_{k=0}^\infty \frac{1}{k!} \left(\frac{i\hbar}{2}\right)^k
    \omega^{\mu_1 \nu_1} \dotsm \omega^{\mu_k \nu_k}
    \iftwocolumn{\nonumber \\ & \quad {} \times}{}
    (\underbrace{\tilde{\nabla} \dotsm \tilde{\nabla}}_k f)_{\mu_1\dotsc \mu_k}
    (\underbrace{\tilde{\nabla} \dotsm \tilde{\nabla}}_k g)_{\nu_1\dotsc \nu_k},
\label{eq:18}
\end{align}
since for Riemann normal coordinates both products coincide. The flatness of the
linear connection $\tilde{\nabla}$ guaranties that the star-product
\eqref{eq:18} is associative.

\begin{remark}
The star-product \eqref{eq:17} is also a valid star-product on more general
symplectic manifolds. Let us consider a symplectic manifold $M$ whose Poisson
tensor can be written in the form \eqref{eq:16}. In addition, let us assume that
the first de Rham cohomology class $H^1(M)$ vanishes. This will guarantee the
existence of global natural coordinate systems associated to the star-products
\eqref{eq:17}. On such symplectic manifold $M$ the product \eqref{eq:17} is a
valid star-product, which can also be written in a covariant form \eqref{eq:18}
with an appropriate linear connection $\tilde{\nabla}$. However, in this case
there is no distinguished star-product from the family of products
\eqref{eq:17}. To distinguish a star-product we have to distinguish a sequence
of commuting vector fields $X_i,Y_i$ from the decomposition \eqref{eq:16} of the
Poisson tensor, or equivalently, by distinguishing a flat torsionless symplectic
linear connection $\tilde{\nabla}$ on $M$.
\end{remark}

An important property of the star-product \eqref{eq:18} used in quantum
mechanics (see \secref{sec:6}) is the fact that for a given classical and
quantum canonical coordinate system $(x,p)$ the star-product \eqref{eq:18}
is equivalent with a Moyal product associated to the coordinates $(x,p)$
(for details and a definition of a quantum canonical coordinate system see
\cite{Blaszak:2013,Blaszak:2013b})
\begin{equation}
f \star_M^{(x,p)} g = f \exp \left(
    \frac{1}{2} i\hbar \overleftarrow{\partial_{x^i}}
    \overrightarrow{\partial_{p_i}}
    - \frac{1}{2} i\hbar \overleftarrow{\partial_{p_i}}
    \overrightarrow{\partial_{x^i}} \right) g.
\label{eq:41}
\end{equation}
In other words there exists a formal series of operators
\begin{equation*}
S = \id + \sum_{k=1}^\infty S_k
\end{equation*}
such that
\begin{equation*}
S(f \star_M^{(x,p)} g) = Sf \star^{(x,p)} Sg.
\end{equation*}
A procedure of a systematic construction of such morphisms can be found in
\cite{Domanski:2013}. Using the results presented in this paper we can construct
the morphism $S$ order by order in $\hbar$. Let us derive the form of $S$ to the
second order in $\hbar$. It happens that only terms with even powers in $\hbar$
are non-zero, thus we only have to calculate $S_2$. To find the form of $S_2$ we
have to solve the following system of equations
\begin{subequations}
\label{eq:26}
\begin{align}
[S_2,z^\alpha] & = -\frac{1}{4}A^\alpha_2, \label{eq:26a} \\
[S_2,\partial^\alpha] & = -\frac{1}{4}A^\alpha_3, \label{eq:26b}
\end{align}
\end{subequations}
where
\begin{align}
A^\alpha_k f & = \frac{1}{k!} \omega^{\mu_1 \nu_1} \dotsm \omega^{\mu_k \nu_k}
    (\underbrace{\tilde{\nabla} \dotsm \tilde{\nabla}}_k
    z^\alpha)_{\mu_1\dotsc\mu_k} \iftwocolumn{\nonumber \\ & \quad {} \times}{}
    (\underbrace{\tilde{\nabla} \dotsm \tilde{\nabla}}_k f)_{\nu_1\dotsc\nu_k},
\label{eq:27}
\end{align}
and $z^i = x^i$, $z^{i+N} = p_i$ for $i = 1,\dotsc,N$,
$\partial^\alpha = \omega^{\alpha \beta} \partial_\beta$.

In what follows we will show that the solution to \eqref{eq:26} is of the form
\begin{equation}
S_2 = -\frac{1}{24} \tilde{\Gamma}_{\alpha \beta \gamma} \partial^\alpha
    \partial^\beta \partial^\gamma
    + \frac{1}{16} \tilde{\Gamma}^\mu_{\nu \alpha}\tilde{\Gamma}^\nu_{\mu \beta}
    \partial^\alpha \partial^\beta,
\label{eq:25}
\end{equation}
where $\tilde{\Gamma}_{\alpha \beta \gamma} = \omega_{\alpha \delta}
\tilde{\Gamma}^\delta_{\beta \gamma}$ (see Appendix for the proof). Note that
the condition that $\tilde{\nabla}$ has vanishing torsion can be restated as
\begin{equation}
\tilde{\Gamma}^\alpha_{\beta \gamma} = \tilde{\Gamma}^\alpha_{\gamma \beta},
\label{eq:23}
\end{equation}
and the condition that $\tilde{\nabla}$ is symplectic ($\omega_{\mu \nu;\alpha}
= 0$, $\omega^{\mu \nu}_{\phantom{\mu \nu};\alpha} = 0$) in Darboux coordinates
can be restated as
\begin{subequations}
\label{eq:24}
\begin{align}
\omega^{\delta \beta} \tilde{\Gamma}^\alpha_{\beta \gamma} & =
    \omega^{\alpha \beta} \tilde{\Gamma}^\delta_{\beta \gamma},
\label{eq:24a} \\
\omega_{\delta \alpha} \tilde{\Gamma}^\alpha_{\beta \gamma} & =
    \omega_{\beta \alpha} \tilde{\Gamma}^\alpha_{\delta \gamma}.
\label{eq:24b}
\end{align}
\end{subequations}
From conditions \eqref{eq:23} and \eqref{eq:24b} we get that $\tilde{\nabla}$ is
symplectic and torsionless iff $\tilde{\Gamma}_{\alpha \beta \gamma}$ is
symmetric with respect to indices $\alpha,\beta,\gamma$ \cite{Plebanski:2001}.

\section{The case of a symplectic manifold $T^*\mathcal{Q}$ with a non-flat base
manifold $\mathcal{Q}$}
\label{sec:5}
In this section we will describe a procedure of introducing star-products on
a symplectic manifold $M = T^*\mathcal{Q}$ over a pseudo-Riemannian manifold
$(\mathcal{Q},g)$ with a Levi-Civita connection induced by a non-flat metric
tensor $g$, where $\mathcal{Q}$ is not necessarily flat and for which does not
necessarily exist a global Riemann normal coordinate system. In such case it is
not possible to introduce a star-product by the formula \eqref{eq:17}, and even
if there would exist global Riemann normal coordinate systems on $\mathcal{Q}$
they would not be related by the formula \eqref{eq:35}, and because of this
different Riemann normal coordinate systems would define different star-products
of the form \eqref{eq:17}.

Henceforth, in such general case we will use a connection $\tilde{\nabla}$ on
$T^*\mathcal{Q}$, induced from a Levi-Civita connection $\nabla$ on
$\mathcal{Q}$, to define a star-product. However, a star-product in the form
\eqref{eq:18} for a curved linear connection $\tilde{\nabla}$ is not a proper
star-product (it is not associative). Thus we have to change the star-product
\eqref{eq:18} in such a way that for a curved linear connection $\tilde{\nabla}$
it would remain associative. Moreover, we would like it to be equivalent with
the Moyal product for every classical and quantum canonical coordinate system.

As a special case we can consider a symplectic manifold $T^*E^N$ with a non-flat
symplectic connection \eqref{eq:33}, \eqref{eq:34} induced by a non-flat
connection defined on $E^N$ (possibly by some non-flat metric). Although in this
case there is a global coordinate chart, the star-product of the form
\eqref{eq:18} is not admissible as well.

The general way of defining on a symplectic manifold $M$ a star-product
equivalent with the Moyal product is as follows.
As in the general case there is no single global coordinate
chart, in order to define a product, which will be equivalent with the Moyal
product, it is necessary to do this locally for every classical and quantum
canonical coordinate chart. Let us take an atlas of classical and quantum
canonical coordinate charts $(x_\alpha,p_\alpha)$ defined on open subsets
$U_\alpha$ of the symplectic manifold $M$. Moreover, let us take some family of
linear automorphisms $S_\alpha$ of $C^\infty(U_\alpha)$ with the property: two
morphisms $S_\alpha$ and $S_\beta$ when acted on the Moyal products
$\star_M^{(x_\alpha,p_\alpha)}$ and $\star_M^{(x_\beta,p_\beta)}$ give
star-products, which on the intersection $U_\alpha \cap U_\beta$, are related to
each other by the change of variables $(x_\alpha,p_\alpha) \mapsto
(x_\beta,p_\beta)$. Every such automorphism $S_\alpha$ can be used to define
a star-product on $C^\infty(U_\alpha)$ by acting on the Moyal product
$\star_M^{(x_\alpha,p_\alpha)}$. All these star-products are consistent on the
intersections $U_\alpha \cap U_\beta$ and hence glue together to give a global
star-product on $C^\infty(M)$. The question whether such family of automorphisms
$S_\alpha$ always exists is nontrivial. Moreover, in the case when such family
exists it is not specified uniquely.

In what follows we will show a way of defining a natural star-product on a
symplectic manifold $M = T^*\mathcal{Q}$ endowed with a non-flat symplectic
torsionless linear connection $\tilde{\nabla}$ induced by a Levi-Civita
connection $\nabla$ on $\mathcal{Q}$. We will present the construction to the
third order in $\hbar$. Let us take the admissible morphisms $S$ ($S_\alpha$) in
the similar form as for the flat case (see formula \eqref{eq:25})
\begin{widetext}
\begin{equation}
S = \id + \hbar^2 \left( -\frac{1}{24} \tilde{\Gamma}_{\alpha \beta \gamma}
    \partial^\alpha \partial^\beta \partial^\gamma
    + \frac{1}{16}(\tilde{\Gamma}^\mu_{\nu \alpha}\tilde{\Gamma}^\nu_{\mu \beta}
    + a\tilde{R}_{\alpha \beta}) \partial^\alpha \partial^\beta \right)
    + o(\hbar^4),
\label{eq:36}
\end{equation}
where $a$ is some real parameter and $\tilde{R}_{\alpha \beta}$ is the Ricci
curvature tensor. Then we will receive the one-parameter family of star-products
in the form
\begin{equation}
f \star_a g = \sum_{k=0}^\infty \frac{1}{k!} \left(\frac{i\hbar}{2}\right)^k
    \omega^{\mu_1 \nu_1} \dotsm \omega^{\mu_k \nu_k} \Bigl(
    (\underbrace{\tilde{\nabla} \dotsm \tilde{\nabla}}_k f)_{\mu_1\dotsc \mu_k}
    (\underbrace{\tilde{\nabla} \dotsm \tilde{\nabla}}_k g)_{\nu_1\dotsc \nu_k}
    + B_{\mu_1 \dotsc \mu_k\nu_1 \dotsc \nu_k}(f,g) \Bigr),
\label{eq:29}
\end{equation}
where $B_{\mu_1 \dotsc \mu_k\nu_1 \dotsc \nu_k}$ are bilinear operators given by
\begin{subequations}
\label{eq:39}
\begin{align}
B_0(f,g) & = 0, \label{eq:39a} \\
B_{\mu_1 \nu_1}(f,g) & = 0, \label{eq:39b} \\
B_{\mu_1 \mu_2 \nu_1 \nu_2}(f,g) & = -a\tilde{R}_{\mu_1 \mu_2}
    (\tilde{\nabla}_{\nu_1}f)(\tilde{\nabla}_{\nu_2}g), \label{eq:39c} \\
B_{\mu_1 \mu_2 \mu_3 \nu_1 \nu_2 \nu_3}(f,g) & =
    -\tilde{R}_{\nu_1 \nu_2 \nu_3 \alpha} \omega^{\alpha \beta}
    (\tilde{\nabla} \tilde{\nabla} \tilde{\nabla} f)_{\mu_1 \mu_2 \mu_3}
    (\tilde{\nabla}_\beta g)
    - \tilde{R}_{\mu_1 \mu_2 \mu_3 \alpha} \omega^{\alpha \beta}
    (\tilde{\nabla}_\beta f)
    (\tilde{\nabla} \tilde{\nabla} \tilde{\nabla} g)_{\nu_1 \nu_2 \nu_3}
    \nonumber \\
& \quad {} - \frac{3}{2}a\tilde{R}_{\mu_1 \mu_2; \mu_3}(\tilde{\nabla}_{\nu_3}f)
    (\tilde{\nabla}\tilde{\nabla}g)_{\nu_1 \nu_2}
    + \frac{3}{2}a\tilde{R}_{\mu_1 \mu_2; \mu_3}
    (\tilde{\nabla}\tilde{\nabla}f)_{\nu_1 \nu_2}
    (\tilde{\nabla}_{\nu_3}g) \nonumber \\
& \quad {} + 3a\tilde{R}_{\mu_2 \nu_3}
    (\tilde{\nabla}\tilde{\nabla}f)_{\mu_1 \mu_3}
    (\tilde{\nabla}\tilde{\nabla}g)_{\nu_1 \nu_2}
    + \tilde{R}_{\mu_1 \mu_2 \mu_3 \alpha}
    \tilde{R}_{\nu_1 \nu_2 \nu_3 \gamma} \omega^{\alpha \beta}
    \omega^{\gamma \delta} (\tilde{\nabla}_\beta f) (\tilde{\nabla}_\delta g),
    \label{eq:39d}
\end{align}
\end{subequations}
\end{widetext}
and $\tilde{R}_{\alpha \beta \gamma \delta} = \omega_{\alpha \lambda}
\tilde{R}^\lambda_{\beta \gamma \delta}$ is the curvature tensor. Analogical
considerations as in the previous section prove that the star-products
\eqref{eq:29} with the four first operators
$B_{\mu_1 \dotsc \mu_k\nu_1 \dotsc \nu_k}$ given by \eqref{eq:39} are equivalent
with the Moyal product, up to third order in $\hbar$. Clearly for the flat
linear connection $\tilde{\nabla}$ the products \eqref{eq:29} reduce to
\eqref{eq:18}.

In a special case $a = 0$ the star-product \eqref{eq:29} reduces to
\begin{equation}
f \star g = \sum_{k=0}^\infty \frac{1}{k!} \left(\frac{i\hbar}{2}\right)^k
    \omega^{\mu_1 \nu_1} \dotsm \omega^{\mu_k \nu_k}
    (D_{\mu_1 \dotsc \mu_k}f)(D_{\nu_1 \dotsc \nu_k}g),
\label{eq:42}
\end{equation}
where $D_{\mu_1 \dotsc \mu_k}$ are linear operators mapping functions to
\nbr{k}times covariant tensor fields given by
\begin{subequations}
\label{eq:30}
\begin{align}
D_0f & = f, \label{eq:30a} \\
D_{\mu_1}f & = \tilde{\nabla}_{\mu_1}f, \label{eq:30b} \\
D_{\mu_1 \mu_2}f & = (\tilde{\nabla}\tilde{\nabla}f)_{\mu_1 \mu_2},
\label{eq:30c} \\
D_{\mu_1 \mu_2 \mu_3}f & =
    (\tilde{\nabla}\tilde{\nabla}\tilde{\nabla}f)_{\mu_1 \mu_2 \mu_3}
    - \tilde{R}_{\mu_1 \mu_2 \mu_3 \alpha} \omega^{\alpha \beta}
    \tilde{\nabla}_\beta f.
\label{eq:30d}
\end{align}
\end{subequations}
A simple calculation, with the help of the Ricci identity
\begin{equation*}
\tilde{R}_{\alpha \beta \gamma \delta} + \tilde{R}_{\alpha \gamma \delta \beta}
    + \tilde{R}_{\alpha \delta \beta \gamma} = 0,
\end{equation*}
shows that operators \eqref{eq:30} are symmetric with respect to indices
$\mu_1,\mu_2,\dotsc$. It is remarkable that the star-product \eqref{eq:42} up to
at least third order in $\hbar$ is a Fedosov star-product associated with the
Weyl curvature form $\Omega = \omega$ \cite{Fedosov:1994}. Whether the Fedosov
star-product for any order in $\hbar$ is of the form \eqref{eq:42} with
operators $D_{\mu_1 \dotsc \mu_k}$ independent on $\hbar$ is an open question
and would be an interesting problem to investigate. It should be noted that for
$a \neq 0$ the star-product \eqref{eq:29} is not a Fedosov star-product.

From the presented construction it is clear that when the configuration space
$\mathcal{Q}$ is curved there is no single natural star-product on
$T^*\mathcal{Q}$ but the whole family of natural star-products. In the
considered case (see formula \eqref{eq:36}) the natural star-products are
parametrized by a real number $a$. Also the Fedosov construction of
star-products has freedom in taking different Weyl curvature forms $\Omega$.

\begin{remark}
The presented construction of the star-products on a symplectic manifold
$T^*\mathcal{Q}$ can be generalized, in a straightforward way, to a general
symplectic manifold $M$ endowed with a symplectic torsionless linear connection
$\tilde{\nabla}$. Formulas \eqref{eq:36}, \eqref{eq:29} and \eqref{eq:39} remain
the same.
\end{remark}

Using \eqref{eq:33} the formula \eqref{eq:36} can be rewritten in the form
\begin{align}
S & = \id + \frac{\hbar^2}{4!} \Bigl(3\left(\Gamma^i_{lj}(x)\Gamma^l_{ik}(x)
    + a R_{jk}(x)\right)\partial_{p_j}\partial_{p_k}
    \iftwocolumn{\nonumber \\ & \quad {}}{}
    + 3\Gamma^i_{jk}(x)\partial_{x^i}\partial_{p_j}\partial_{p_k}
    \nonumber \\
& \quad {} + \left( 2\Gamma^i_{nl}(x)\Gamma^n_{jk}(x)
    - \partial_{x^l}\Gamma^i_{jk}(x) \right)
    p_i \partial_{p_j}\partial_{p_k}\partial_{p_l} \Bigr)
    \iftwocolumn{\nonumber \\ & \quad {}}{}
    + o(\hbar^4).
\label{eq:37}
\end{align}
Let us generalize the formula \eqref{eq:37} in the following way
\begin{align}
S & = \id + \frac{\hbar^2}{4!} \Bigl(3\left(\Gamma^i_{lj}(x)\Gamma^l_{ik}(x)
    + a R_{jk}(x)\right)\partial_{p_j}\partial_{p_k}
    \iftwocolumn{\nonumber \\ & \quad {}}{}
    + 3\Gamma^i_{jk}(x)\partial_{x^i}\partial_{p_j}\partial_{p_k}
    \nonumber \\
& \quad {} + \left( 2\Gamma^i_{nl}(x)\Gamma^n_{jk}(x)
    - \partial_{x^l}\Gamma^i_{jk}(x) \right)
    p_i \partial_{p_j}\partial_{p_k}\partial_{p_l} \nonumber \\
& \quad {} - 3b\partial_{p_j}(\partial_{x^j} + \Gamma^i_{jl} p_i \partial_{p_l})
    \partial_{p_k}(\partial_{x^k} + \Gamma^r_{kn} p_r \partial_{p_n}) \Bigr)
    \iftwocolumn{\nonumber \\ & \quad {}}{}
    + o(\hbar^4),
\label{eq:38}
\end{align}
where $b$ is some real parameter. The star-product induced by the above morphism
$S$ for $a = 1$ and $b = 1$ leads to what was called in a paper
\cite{Duval:2005} a ``minimal'' quantization. Moreover, the same quantization
was used in \cite{Benenti:2002a,Benenti:2002b,Blaszak:2013c} in order to
investigate the quantum integrability and quantum separability of classical
St\"ackel systems.

\section{Quantum mechanical operators}
\label{sec:6}
To star-algebras $(C^\infty(M),\star)$ are associated algebras of operators
defined on certain Hilbert spaces. In \cite{Blaszak:2013,Blaszak:2013b} was
presented a construction of such algebras of operators for a given classical and
quantum canonical coordinate system. In this section we will use the results
from \cite{Blaszak:2013,Blaszak:2013b} to derive a construction of such algebras
of operators in a coordinate independent way.

We will be considering a symplectic manifold $M = T^*\mathcal{Q}$ over a
pseudo-Riemannian manifold $(\mathcal{Q},g)$, and a family of star-products
on $M$ considered in \secref{sec:5}. Let us introduce the notion of an
almost global coordinate system. The coordinate system
$\phi \colon \mathcal{Q} \supset U \to V \subset \mathbb{R}^N$ is called an
almost global coordinate system on $\mathcal{Q}$ if $\mathcal{Q} \setminus U$
is of measure zero with respect to a measure given by the metric volume form
$\omega_g$. Similarly we define an almost global coordinate system on
$T^*\mathcal{Q}$ where as a measure on $T^*\mathcal{Q}$ we take a measure
induced by a Liouville form
\begin{equation*}
\Omega = \frac{1}{N!}\underbrace{\omega \wedge \dotsm \wedge \omega}_N.
\end{equation*}
A Darboux coordinate system induced from
an almost global coordinate system on $\mathcal{Q}$ is the almost global
coordinate system on $T^*\mathcal{Q}$. In what follows we will consider only
spaces $\mathcal{Q}$ which admit an almost global coordinate system.

Let us consider a Hilbert space $L^2(M,\Omega)$ of functions defined on the
symplectic manifold $M = T^*\mathcal{Q}$, square integrable with respect to the
Liouville form $\Omega$. Let us also consider a Hilbert space
$L^2(\mathcal{Q},\omega_g)$ of functions defined on $\mathcal{Q}$ and square
integrable with respect to the metric volume form $\omega_g$. To every
$A \in C^\infty(M)$ we can associate an operator $\hat{A}$, defined on the
Hilbert space $L^2(M,\Omega)$, by the formula
\begin{equation*}
\hat{A}\Psi = A \star \Psi,
\end{equation*}
for every smooth $\Psi \in L^2(M,\Omega)$. To function $A$ we can also associate
an operator defined on the Hilbert space $L^2(\mathcal{Q},\omega_g)$. To
construct such operator first let us consider an almost global coordinate system
on $\mathcal{Q}$, $\phi\colon \mathcal{Q} \supset U \to V \subset \mathbb{R}^N$,
and related to it an almost global classical and quantum canonical coordinate
system on $M$, $\tilde{\phi} \colon M \supset \mathcal{U} \to \mathcal{V}
\subset \mathbb{R}^{2N}$. The coordinate system $\phi$ defines a natural
isomorphism $F_\phi \colon L^2(\mathcal{Q},\omega_g) \to L^2(V,\mu)$ between the
Hilbert space $L^2(\mathcal{Q},\omega_g)$ and a Hilbert space $L^2(V,\mu)$,
where $\dd{\mu(x)} = \abs{\det[g_{ij}(x)]}^{1/2} \ud{x}$:
\begin{equation*}
F_\phi \psi = \psi|_U \circ \phi^{-1}.
\end{equation*}
Similarly, the coordinate system $\tilde{\phi}$ defines a natural isomorphism
$\tilde{F}_{\tilde{\phi}} \colon L^2(M,\Omega) \to L^2(\mathcal{V})$ between the
Hilbert space $L^2(M,\Omega)$ and a Hilbert space $L^2(\mathcal{V})$ of
functions defined on $\mathcal{V}$ and square integrable with respect to the
Lebesgue measure:
\begin{equation*}
\tilde{F}_{\tilde{\phi}} \Psi = \Psi|_{\mathcal{U}} \circ \tilde{\phi}^{-1}.
\end{equation*}

According to \cite{Blaszak:2013b} the Hilbert space $L^2(\mathcal{V})$ can be
written as the following tensor product
\begin{align*}
L^2(\mathcal{V}) & = (L^2(V,\mu))^* \otimes_S L^2(V,\mu) \iftwocolumn{\\ & }{}
= S\bigl((L^2(V,\mu))^* \otimes_M L^2(V,\mu)\bigr),
\end{align*}
where $S$ is the morphism \eqref{eq:38} intertwining the
\nbr{\star^{(x,p)}}product with the Moyal product $\star_M^{(x,p)}$,
$(L^2(V,\mu))^*$ is the dual space to $L^2(V,\mu)$, and $\otimes_M$ is a
Wigner-Moyal transform \cite{Gosson:2005}. Using the isomorphisms
$F_\phi$ and $\tilde{F}_{\tilde{\phi}}$ we can write $L^2(M,\Omega)$ as the
following tensor product
\begin{equation*}
L^2(M,\Omega) = (L^2(\mathcal{Q},\omega_g))^* \otimes
    L^2(\mathcal{Q},\omega_g),
\end{equation*}
where
\begin{equation*}
\varphi^* \otimes \psi = (F_\phi^{-1})^*\varphi^* \otimes_S F_\phi \psi,
\quad \varphi,\psi \in L^2(\mathcal{Q},\omega_g).
\end{equation*}
Note that the above definition of the tensor product $\otimes$ is independent of
the choice of the coordinate system $\phi$. Moreover, to an operator
$A \star^{(x,p)} {}$, where $A \in C^\infty(\mathcal{V})$, we can associate an
\nbr{S}ordered operator $A_S(\hat{q},\hat{p})$ by the formula
\cite{Blaszak:2013b}
\begin{equation*}
A \star^{(x,p)} {} = \hat{1} \otimes_S A_S(\hat{q},\hat{p}),
\end{equation*}
where
\begin{equation*}
A_S(\hat{q},\hat{p})=(S^{-1}A)_W(\hat{q},\hat{p}),
\end{equation*}
$S$ relates star-product $\star^{(x,p)}$ with Moyal product $\star^{(x,p)}_M$
and $W$ means the Weyl (symmetric) ordering of operators
$\hat{q}^i$, $\hat{p}_j$, which are canonical operators of
position and momentum associated to  the Levi-Civita connection
$\nabla$ in the coordinate system $\tilde{\phi}$:
\begin{align*}
\hat{q}^i & = x^i, \\
\hat{p}_j & = -i\hbar\left(\partial_{x^j} + \frac{1}{2}\Gamma^k_{jk}(x)\right).
\end{align*}
Again, using the isomorphisms $F_\phi$ and $\tilde{F}_{\tilde{\phi}}$, we can
see that to every operator $A \star {}$, where $A \in C^\infty(M)$, we can
associate an operator $\hat{A}$, defined on the Hilbert space
$L^2(\mathcal{Q},\omega_g)$, by the formula
\begin{equation*}
A \star {} = \hat{1} \otimes \hat{A}.
\end{equation*}
The operator $\hat{A}$ has the property that for any almost global coordinate
system on $\mathcal{Q}$ it takes the form of an \nbr{S}ordered operator
$A_S(\hat{q},\hat{p})$.

In what follows let us give examples of operators, defined on the
Hilbert space $L^2(\mathcal{Q},\omega_g)$ and written in an
invariant form, associated to functions (observables) linear,
quadratic and cubic in momenta. The derivation of the formulas
presented below is analogous as in \cite{Blaszak:2013b}. The
connection $\nabla$ is fixed by $g$ and an appropriate
$\star$ (quantization) is chosen by fixing a particular $S$
\eqref{eq:38}. Let $H$ be a function on $M$ which in some Darboux
coordinate system $(x,p)$ takes the form
\begin{equation*}
H(x,p) = K^i(x)p_i,
\end{equation*}
where $K^i(x)$ are components of some vector field $K$ on
$\mathcal{Q}$. To the function $H$ corresponds the following
hermitian operator $\hat{H}$ in $L^2(\mathcal{Q},\omega_g)$
\begin{equation*}
\hat{H} = -\frac{i\hbar}{2}\left(K^i \nabla_i + \nabla_i K^i\right).
\end{equation*}
Similarly, let now $H$ be a function on $M$ which in $(x,p)$ coordinates takes
the form
\begin{equation*}
H(x,p) = K^{ij}(x) p_i p_j,
\end{equation*}
where $K^{ij}(x)$ are components of some symmetric second order
tensor field $K$ on $\mathcal{Q}$. To the function $H$ corresponds
the hermitian operator
\begin{align*}
\hat{H} & = -\hbar^2 \biggl(\nabla_i K^{ij} \nabla_j
    + \frac{1}{4}(1 - b) K^{ij}_{\phantom{ij};ij}
    \iftwocolumn{\nonumber \\ & \quad {}}{}
    - \frac{1}{4}(1 - a) K^{ij} R_{ij} \biggr),
\end{align*}
where $;i$ denotes the covariant derivative in the direction of
the vector field $\partial_{x^i}$. Finally, let $H$ be a function
on $M$ which in $(x,p)$ coordinates takes the form
\begin{equation*}
H(x,p) = K^{ijk}(x) p_i p_j p_k,
\end{equation*}
where $K^{ijk}(x)$ are components of some symmetric third order
tensor field $K$ on $\mathcal{Q}$. To the function $H$ corresponds
the respective hermitian operator
\begin{align*}
\hat{H} & = \frac{1}{2} i\hbar^3 \biggl(
    \nabla_i K^{ijk} \nabla_j \nabla_k + \nabla_i \nabla_j K^{ijk} \nabla_k
    \iftwocolumn{\nonumber \\ & \quad {}}{}
    + \frac{1}{4}(1 - b) \nabla_k K^{ijk}_{\phantom{ijk};ij}
    + \frac{1}{4}(1 - b) K^{ijk}_{\phantom{ijk};ij} \nabla_k \nonumber \\
& \quad {} - \frac{3}{4}(1 - a) \nabla_i K^{ijk} R_{jk}
    - \frac{3}{4}(1 - a) K^{ijk} R_{jk} \nabla_i \biggr).
\end{align*}

Observe that for flat connections we deal with a one parameter ($b$) family of
admissible quantizations. By admissible we understand these quantizations which
coincide for a class of `natural' Hamiltonians
\begin{equation}
H(x,p) = \frac{1}{2} g^{ij}(x) p_i p_j + V(x).
\label{eq:40}
\end{equation}
Notice also that the well known Weyl quantization, written in a coordinate free
form, is the one with $b = 0$. The case $b = 1$ represents a so called flat
minimal quantization. Then, for non-flat connections, we introduced a two
parameter $(a,b)$ family of quantizations which particular representatives the
reader can find in the literature
\cite{DeWitt:1957,Dekker:1980,Liu:1992,Duval:2005}. The quantizations
$(a,b) = (a,0)$ represent non-flat generalizations of Weyl quantization, while
the case $(a,b) = (1,1)$ is mentioned previously non-flat minimal quantization.

\section{Final remarks}
\label{sec:7}
In this paper we investigated a problem of defining natural star-products on
symplectic manifolds $M = T^*\mathcal{Q}$. We also associated to considered
star-algebras operator algebras defined on certain Hilbert spaces. All this is
the main ingredient of a quantization procedure of classical Hamiltonian systems
\cite{Blaszak:2013b}. Thus the first step in quantizing a classical Hamiltonian
system is to \nbr{\hbar}deform a classical Poisson algebra $C^\infty(M)$ to a
quantum Poisson algebra $(C^\infty(M;\hbar),\star)$. The presented construction
of the star-products depended on the linear connection $\nabla$ on
$\mathcal{Q}$. Thus the quantization is partly fixed by fixing a linear
connection on $\mathcal{Q}$. However, this does not fix the quantization
entirely as was seen in \secref{sec:5} where we introduced \nbr{(a,b)}parameter
family of star-products for a given linear connection $\nabla$.

Moreover, we have a freedom in choosing quantum observables. Thus the second
step in quantizing a classical Hamiltonian system is to \nbr{\hbar}deform
classical observables $A_C \in C^\infty(M)$, in particular Hamiltonian functions
$H$, to quantum observables $A_Q \in C^\infty(M;\hbar)$. However, the choice of
a star-product, for a given linear connection $\nabla$, and the choice of
quantum observables is somewhat connected. If we choose two star-products
$\star$ and $\star'$, such that there exists a morphism $S$ intertwining these
two star-products, and if we choose two algebras of quantum observables in a way
that they also will be related by the morphism $S$, then such two quantizations
will be equivalent.

As an example let us consider a star-product \eqref{eq:29} written in some local
coordinate system. Instead of using this extremely complex product and quantum
observables equal to the classical ones: $A_Q = A_C$, it is reasonable to use
the Moyal star-product \eqref{eq:41} in these coordinates and take quantum
observables $A_Q$ as an $S$ deformation of the classical ones
\begin{align*}
A_Q & = S^{-1}A_C \iftwocolumn{\nonumber \\ &}{}
= A_C - \frac{\hbar^2}{4!} \Bigl(3\left(\Gamma^i_{lj}(x)\Gamma^l_{ik}(x)
    + a R_{jk}(x)\right)\partial_{p_j}\partial_{p_k}
    \iftwocolumn{\nonumber \\ & \quad {}}{}
    + 3\Gamma^i_{jk}(x)\partial_{x^i}\partial_{p_j}\partial_{p_k}
    \nonumber \\
& \quad {} + \left( 2\Gamma^i_{nl}(x)\Gamma^n_{jk}(x)
    - \partial_{x^l}\Gamma^i_{jk}(x) \right)
    p_i \partial_{p_j}\partial_{p_k}\partial_{p_l} \Bigr)A_C
    \iftwocolumn{\nonumber \\ & \quad {}}{}
    + o(\hbar^4),
\end{align*}
where the morphism $S$ \eqref{eq:37} relates the star-product \eqref{eq:29} with
the Moyal one \eqref{eq:41}.

Hence, an explicit choice of quantization of a classical Hamiltonian system is
fixed by a choice of a linear connection $\nabla$ on $\mathcal{Q}$, and
a star-product related to $\nabla$ (or just the morphism $S$ relating this
star-product with the Moyal one). The choice of the linear connection $\nabla$
on the configuration space $\mathcal{Q}$ is dictated by the classical system
being quantized. For example, if a Hamiltonian of the system is in the natural
form \eqref{eq:40} then the only natural choice is the Levi-Civita connection.
However, if the Hamiltonian of a system is of the form
\begin{equation*}
H(x,p) = \frac{1}{2}K^{ij}p_i p_j + V(x)
= \frac{1}{2}K^i_r g^{rj} p_i p_j + V(x),
\end{equation*}
where $K$ is some symmetric non-degenerate tensor, then we have two different
natural choices of a connection. One is again the Levi-Civita connection induced
by $g$ and the second one is the connection induced by a new metric
$\tilde{g} = K$.

Thus, there is a freedom in choosing a quantization of a given Hamiltonian
system. Only in limited cases we can verify through experiment which
quantization scheme realizes in nature.

It should be noted that we considered quantization of systems over a phase space
$T^*\mathcal{Q}$. For this special case of a phase space it was possible to
introduce an operator representation of the quantum system in a Hilbert space
$L^2(\mathcal{Q},\omega_g)$. The quantization procedure described in the paper
can be generalized to systems defined over a phase space $M$ being a general
symplectic manifold, provided that we fix on $M$ a symplectic torsionless linear
connection. In this case, however, it is difficult to introduce an operator
representation in a Hilbert space being an analog of
$L^2(\mathcal{Q},\omega_g)$.

\begin{widetext}
\section*{Appendix}
\renewcommand{\theequation}{A.\arabic{equation}}
Let us check if $S_2$ in the form \eqref{eq:25} satisfies the system of
equations \eqref{eq:26}. From \eqref{eq:27} and \eqref{eq:24a} we get that
\begin{equation*}
A^\alpha_2 = -\frac{1}{2}\omega^{\mu_1 \nu_1} \omega^{\mu_2 \nu_2}
    \tilde{\Gamma}^\alpha_{\mu_1 \mu_2}(\partial_{\nu_1} \partial_{\nu_2}
    - \tilde{\Gamma}^\beta_{\nu_1 \nu_2} \partial_\beta)
= -\frac{1}{2} \tilde{\Gamma}^\alpha_{\mu_1 \mu_2} \partial^{\mu_1}
    \partial^{\mu_2}
    - \frac{1}{2} \omega^{\mu_1 \alpha} \tilde{\Gamma}^{\nu_1}_{\mu_1 \mu_2}
    \tilde{\Gamma}^{\mu_2}_{\nu_1 \nu_2} \partial^{\nu_2}.
\end{equation*}
On the other hand
\begin{align*}
[S_2,z^\alpha] & = -\frac{1}{24} \omega^{\delta \alpha}
    \tilde{\Gamma}_{\delta\beta\gamma} \partial^\beta \partial^\gamma
    - \frac{1}{24} \omega^{\beta \alpha} \tilde{\Gamma}_{\delta \beta \gamma}
    \partial^\delta \partial^\gamma
    - \frac{1}{24} \omega^{\gamma \alpha} \tilde{\Gamma}_{\delta \beta \gamma}
    \partial^\delta \partial^\beta
    + \frac{1}{16} \omega^{\gamma \alpha} \tilde{\Gamma}^\mu_{\nu \gamma}
    \tilde{\Gamma}^\nu_{\mu \beta} \partial^\beta
    + \frac{1}{16} \omega^{\beta \alpha} \tilde{\Gamma}^\mu_{\nu \gamma}
    \tilde{\Gamma}^\nu_{\mu \beta} \partial^\gamma \\
& = \frac{1}{8} \tilde{\Gamma}^\alpha_{\beta
\gamma}\partial^{\beta}
    \partial^{\gamma}
    + \frac{1}{8} \omega^{\gamma \alpha} \tilde{\Gamma}^{\mu}_{\nu \gamma}
    \tilde{\Gamma}^{\nu}_{\mu \beta} \partial^{\beta},
\end{align*}
which proves \eqref{eq:26a}. From \eqref{eq:27} we can calculate that
\begin{align*}
A^\alpha_3 & = \frac{1}{6} \omega^{\mu_1\nu_1}\omega^{\mu_2\nu_2}
    \omega^{\mu_3\nu_3}
    (\tilde{\nabla}\tilde{\nabla}\tilde{\nabla}z^\alpha)_{\mu_1 \mu_2 \mu_3}
    \Bigl(\partial_{\nu_1} \partial_{\nu_2} \partial_{\nu_3}
    - \tilde{\Gamma}^\beta_{\nu_1 \nu_2} \partial_{\nu_3} \partial_{\beta}
    - \tilde{\Gamma}^\beta_{\nu_3 \nu_1} \partial_{\nu_2} \partial_{\beta}
    - \tilde{\Gamma}^\beta_{\nu_2 \nu_3} \partial_{\nu_1} \partial_{\beta} \\
& \quad{}+(\tilde{\nabla}\tilde{\nabla}\tilde{\nabla}z^\beta)_{\nu_1\nu_2\nu_3}
    \partial_\beta\Bigr).
\end{align*}
The above equation can be rewritten in a different form. To do this first let us
prove that
\begin{subequations}
\label{eq:28}
\begin{align}
\omega^{\mu_1 \nu_1}
    (\tilde{\nabla}\tilde{\nabla}\tilde{\nabla}z^\alpha)_{\mu_1 \mu_2 \mu_3} & =
    \omega^{\alpha \mu_1} \tilde{\Gamma}^{\nu_1}_{\mu_2 \mu_3, \mu_1}
    + \omega^{\alpha \mu_1} \tilde{R}^{\nu_1}_{\mu_2 \mu_3 \mu_1},
    \label{eq:28a} \\
\omega^{\mu_2 \nu_2}
    (\tilde{\nabla}\tilde{\nabla}\tilde{\nabla}z^\alpha)_{\mu_1 \mu_2 \mu_3} & =
    \omega^{\alpha \mu_2} \tilde{\Gamma}^{\nu_2}_{\mu_1 \mu_3, \mu_2}
    + \omega^{\alpha \mu_2} \tilde{R}^{\nu_2}_{\mu_1 \mu_3 \mu_2}.
    \label{eq:28b}
\end{align}
\end{subequations}
Indeed, with the help of \eqref{eq:24} we can calculate that
\begin{align*}
\omega^{\mu_1 \nu_1}
    (\tilde{\nabla}\tilde{\nabla}\tilde{\nabla}z^\alpha)_{\mu_1 \mu_2 \mu_3} & =
    \omega^{\mu_1 \nu_1}(-\tilde{\Gamma}^\alpha_{\mu_2 \mu_1, \mu_3}
    + \tilde{\Gamma}^\beta_{\mu_1 \mu_3} \tilde{\Gamma}^\alpha_{\beta \mu_2}
    + \tilde{\Gamma}^\beta_{\mu_2 \mu_3} \tilde{\Gamma}^\alpha_{\beta \mu_1})\\
& = \omega^{\mu_1 \alpha}(-\tilde{\Gamma}^{\nu_1}_{\mu_2 \mu_1,\mu_3}
    + \tilde{\Gamma}^\beta_{\mu_2 \mu_3} \tilde{\Gamma}^{\nu_1}_{\beta \mu_1})
    + \omega^{\mu_1 \beta} \tilde{\Gamma}^{\nu_1}_{\mu_1 \mu_3}
    \tilde{\Gamma}^\alpha_{\beta \mu_2} \\
& = \omega^{\mu_1 \alpha}(R^{\nu_1}_{\mu_2 \mu_1 \mu_3}
    - \tilde{\Gamma}^{\nu_1}_{\mu_2 \mu_1, \mu_3}
    + \tilde{\Gamma}^\beta_{\mu_2 \mu_3} \tilde{\Gamma}^{\nu_1}_{\beta \mu_1})
    + \omega^{\mu_1 \beta} \tilde{\Gamma}^{\nu_1}_{\mu_1 \mu_3}
    \tilde{\Gamma}^\alpha_{\beta \mu_2},
\end{align*}
and that
\begin{align*}
\omega^{\mu_1 \beta} \tilde{\Gamma}^{\nu_1}_{\mu_1 \mu_3}
    \tilde{\Gamma}^\alpha_{\beta \mu_2} & = \omega^{\mu_1 \beta}
    \delta^\alpha_\gamma
    \tilde{\Gamma}^{\nu_1}_{\mu_1 \mu_3} \tilde{\Gamma}^\gamma_{\beta \mu_2}
= \omega^{\mu_1 \beta} \omega^{\alpha \delta} \omega_{\delta \gamma}
    \tilde{\Gamma}^{\nu_1}_{\mu_1 \mu_3} \tilde{\Gamma}^\gamma_{\beta \mu_2}
= \omega^{\mu_1 \beta} \omega^{\alpha \delta} \omega_{\beta \gamma}
    \tilde{\Gamma}^{\nu_1}_{\mu_1 \mu_3} \tilde{\Gamma}^\gamma_{\delta \mu_2} \\
& = \delta^{\mu_1}_\gamma \omega^{\alpha \delta}
    \tilde{\Gamma}^{\nu_1}_{\mu_1 \mu_3} \tilde{\Gamma}^\gamma_{\delta \mu_2}
= -\omega^{\delta \alpha}
    \tilde{\Gamma}^{\nu_1}_{\mu_1 \mu_3} \tilde{\Gamma}^{\mu_1}_{\delta \mu_2},
\end{align*}
from which follows \eqref{eq:28a}. \eqref{eq:28b} can be proved analogically.
Hence using \eqref{eq:24a}, \eqref{eq:28} and the condition
\begin{equation*}
\omega^{\mu_1 \nu_1} \dotsm \omega^{\mu_k \nu_k}
    (\tilde{\nabla} \dotsm \tilde{\nabla} z^\alpha)_{\mu_1\dotsc\mu_k}
    (\tilde{\nabla} \dotsm \tilde{\nabla} z^\beta)_{\nu_1\dotsc\nu_k} = 0, \quad
k = 3,5,\dotsc
\end{equation*}
following from the quantum canonicity of the coordinate system
$(z^1,\dotsc,z^{2N})$ we get
\begin{equation*}
A^\alpha_3 = \frac{1}{6} \omega^{\alpha \mu_1}
    \left(\tilde{\Gamma}^{\nu_1}_{\mu_2 \mu_3,\mu_1}
    + \tilde{R}^{\nu_1}_{\mu_2 \mu_3 \mu_1}\right) \partial_{\nu_1}
    \partial^{\mu_2} \partial^{\mu_3}
    + \frac{1}{2} \omega^{\alpha \mu_1}
    \left(\tilde{\Gamma}^{\nu_1}_{\mu_2\mu_3,\mu_1}
    + \frac{1}{3}\tilde{R}^{\nu_1}_{\mu_2 \mu_3 \mu_1}
    + \frac{2}{3}\tilde{R}^{\nu_1}_{\mu_3 \mu_2 \mu_1}\right)
    \tilde{\Gamma}^{\mu_2}_{\nu_1 \nu_2} \partial^{\mu_3} \partial^{\nu_2}.
\end{equation*}
On the other hand
\begin{equation*}
[S_2,\partial^\alpha] = -\frac{1}{24} \omega^{\alpha \delta}
    \tilde{\Gamma}^\lambda_{\beta \gamma, \delta} \partial_\lambda\partial^\beta
    \partial^\gamma
    - \frac{1}{8} \omega^{\alpha \delta} \tilde{\Gamma}^\nu_{\mu \beta, \delta}
    \tilde{\Gamma}^\mu_{\nu \lambda} \partial^\lambda \partial^\beta,
\end{equation*}
which shows that $S_2$ in the form \eqref{eq:25} will satisfy \eqref{eq:26b}
since from flatness assumption $\tilde{R}^\alpha_{\beta \gamma \delta} = 0$.
\end{widetext}


%

\end{document}